\documentclass[12pt, draftclsnofoot, onecolumn]{IEEEtran}
\IEEEoverridecommandlockouts
\usepackage[letterpaper, left=1in, right=1in, bottom=1in, top=0.75in]{geometry}
\usepackage[dvips]{graphicx}
 \usepackage[bookmarks=false]{hyperref}

\usepackage[cmex10]{amsmath}
\usepackage{array, setspace}
\usepackage{mdwmath}
\usepackage{mdwtab}
\usepackage[usenames,dvipsnames]{color}
\usepackage{cases}
\usepackage{multirow, lscape}
\usepackage[ampersand]{easylist}
\usepackage{graphicx}
\usepackage{color}
\usepackage{cite}
\usepackage{graphicx}
\usepackage{graphicx, subfigure,floatrow}
\usepackage[cmex10]{amsmath}
\usepackage{esint}
\usepackage{amssymb}
\usepackage{cases}
\usepackage{epstopdf}
\usepackage{amsthm}
\usepackage{bm}
\usepackage{caption}
\usepackage{setspace}
\usepackage{stfloats}
\usepackage{verbatim}
\usepackage{doi}
\usepackage{soul}
\sethlcolor{yellow}
\soulregister\cite7 
\soulregister\ref7 
\soulregister\eqref7 

\usepackage{amssymb}
\usepackage{mathptmx}
\usepackage{eucal}
\usepackage{underscore}
\usepackage{subfigure}
\usepackage{array}

\usepackage{amsmath}

\usepackage{ulem}

\usepackage{mathrsfs}
\usepackage[ruled,linesnumberedhidden,longend]{algorithm2e}

\usepackage{cite}

\def\BibTeX{{\rm B\kern-.05em{\sc i\kern-.025em b}\kern-.08em
    T\kern-.1667em\lower.7ex\hbox{E}\kern-.125emX}}

\begin{document}

\title{ \textcolor[rgb]{0,0,0}{A Layered Grouping Random Access Scheme Based on Dynamic Preamble Selection for Massive Machine Type Communications} }

 \author{
  Gaofeng Cheng, Huan Chen, Pingzhi Fan, Li Li, Li Hao  \\

\vspace{-1cm}
\thanks{
Gaofeng Cheng, Huan Chen, Pingzhi Fan, L. Li, and L. Hao are with the School of Information Science and Technology, Southwest Jiaotong University, Chengdu, 610031, China (e-mail: chenggaofeng21@gmail.com; huan.chen0728@foxmail.com; \{pzfan, ll5e08, lhao\}@home.swjtu.edu.cn).
}
}
	

\maketitle

\begin{abstract}

Massive machine type communication (mMTC) has been identified as an important use case in Beyond 5G networks and future massive Internet of Things (IoT). However, for the massive multiple access in mMTC, there is a serious access preamble collision problem if the conventional 4-step random access (RA) scheme is employed. Consequently, a range of grant-free (GF) RA schemes were proposed.
Nevertheless, if the number of \textcolor[rgb]{0,0,0}{cellular} users (devices) significantly increases, both the energy and spectrum efficiency of the existing \textcolor[rgb]{0,0,0}{GF} schemes still rapidly degrade owing to the much longer preambles required. In order to overcome this dilemma, a \textcolor[rgb]{0,0,0}{layered} grouping strategy is proposed, where the \textcolor[rgb]{0,0,0}{cellular} users are firstly divided into clusters based on their geographical locations, and then the users of the same cluster autonomously join in different groups by using optimum energy consumption (Opt-EC) based K-means algorithm.
\textcolor[rgb]{0,0,0}{With this new layered cellular architecture}, the RA process is divided into cluster load estimation phase and active group detection phase. Based on the state evolution theory of approximated message passing algorithm, a tight lower bound on the minimum preamble length for achieving a certain detection accuracy is derived. Benefiting from the cluster load estimation, a dynamic preamble selection (DPS) strategy is invoked in the second phase, resulting the required preambles with minimum length. As evidenced in our simulation results, this \textcolor[rgb]{0,0,0}{two-phase DPS aided RA} strategy results in a significant performance improvement.

\textit{\textbf{Index Terms --- }}\textbf{grant-free, layered grouping, AMP algorithm, minimum preamble length.}
\end{abstract}


\section{Introduction}
\label{Introduction}

In the next generation cellular networks, massive machine-type communications (mMTC) will play an essential role for building the massive Internet of Things (IoT) and hence have been identified as one of the three main use cases in 5G and Beyond 5G (B5G) services. The mMTC service has its particular features:
(1) compared with human-to-human (H2H) communications, the number of potential users (devices) in mMTC scenario could reach up to millions \cite{3GPP1808866};
(2) the user activity patterns are very sporadic; 
(3) mMTC demonstrates a salient short-packet-transmission property \cite{ShortPacketsBockelmann};
(4) the users are very sensitive to energy consumption \cite{3GPP1905187}.

Let $N$ represent the total number of potential users in a cell and $N_{ac}$ the number of active users during one random access opportunity (RAO). Hence, the system sparsity could be defined as $\lambda = \frac{N_{ac}}{N}$. Even $\lambda$ is normally small in mMTC scenario, $N_{ac}$ is still significantly larger than the \textcolor[rgb]{0,0,0}{size of LTE preamble pool}. Because $N$ is extremely large \cite{Durisi2016Toward}.  As a result, the traditional \textcolor[rgb]{0,0,0}{LTE} grant-based four-step random access will encounter a serious preamble collision problem, which dramatically reduces the access success probability and increases the access latency \cite{ShortPacketsBockelmann}. \textcolor[rgb]{0,0,0}{Implementing the synchronization, active user detection, channel estimation, as well as the data recovery in a one-shot joint operation becomes a promising direction, in grant-free (GF) RA \cite{3GPP165174, 3GPP166095, 3GPP164268}. Since GF RA realizes an "arrival-and-go" transmission of payload, it has} attracted significant attentions in recent years. 

\textcolor[rgb]{0,0,0}{Obviously, the non-orthogonal multiple access (NOMA) technology directly motivate the concept of GF RA as NOMA allows multiple users to share the same time-frequency resource. Base Station (BS) can distinguish multi-user data through their different signature patterns if the system overload does not exceed a certain level. Hence the cellular users can transmit their data whenever available. A range of GF NOMA schemes were proposed, including power-based GF NOMA \cite{Ding2014On}, spreading-based GF NOMA \cite{Bayesteh2014Blind, Wanwei2018Uplink, Yuan2017Blind }, interleaving-based GF NOMA \cite{3GPP164269}, etc. However, before grant-free transmission, GF NOMA schemes still require necessary overheads to tackle with the synchronization, identification and channel estimation of active users. Some of the proposals even assume that the active users already connect to BS, or the active users and BS know almost everything about each other, such as the number of multiplexing users, their modulation and coding schemes. These facts imply that realizing an idealized GF RA is extremely challenging, and the required overhead may be inevitably large in practice.} 

\textcolor[rgb]{0,0,0}{As a result, reducing the overhead for GF RA becomes an important issue. In this spirit, compressive sensing (CS) technology becomes another foundation to enable GF RA \cite{Fazel2013Random, Hong2015Sparsity }. Because CS is capable of recovering the desired signals from far fewer measurements than the total signal dimensions if a certain signal sparsity is guaranteed, CS is normally employed in GF RA schemes to overcome the challenges of user identification and channel estimation \cite{Chen2018Sparse, PartI2018}, or even simultaneously recover the payload data \cite{Senel2018Grant}}.

\textcolor[rgb]{0,0,0}{In the CS based GF RA, each device is assigned a user specific pilot sequence, termed as preamble. According to CS theory, the preamble length for enabling a successful signal reconstruction is impacted by the number of total users, the sparsity, and the type of measurement matrix. This preamble length is also regarded as a dominant overhead metric of CS based GF RA. Therefore, investigating the minimum preamble length (MPL) is important to mitigate high overhead problem. The associated theoretical analysis has been attempted in \cite{DecodingTao2005, Donoho2003}. However, only an asymptotic order of MPL was obtained. Some uncertain parameters were still involved in this asymptotic order, whose value have to be experimentally tested according to specific scenarios. This problem obviously limits the application of these theoretical results. On the other hand, with the increasing number of potential users or active users, the preamble length has to be increased accordingly. Consequently, in future ultra-dense cellular IoT networks ($\textgreater 10^6$ devices$/km^2$) \cite{3GPP1903968}, the high overhead problem may still exist even employing CS based GF RA, which limits the number of affordable users within the same cell and also aggravate the constrained power budget of  RA procedure.}

\textcolor[rgb]{0,0,0}{Extending our horizon further}, to address the overload problem in future random access channel (RACH),  a range of other methods have been proposed \cite{Sharma2020Toward}. These existing approaches could be categorized into push-based and pull-based. In push-based approaches, the RA requests are \textcolor[rgb]{0,0,0}{triggered from} the device side while in pull-based approaches, the contention is controlled  from the \textcolor[rgb]{0,0,0}{BS} side. \textcolor[rgb]{0,0,0}{Among these methods, it is noticeable that}, grouping is an efficient alternative to \textcolor[rgb]{0,0,0}{relax the cellular density, hence facilitates the massive connectivity and reduces the energy consumption of RA procedure.} That is, all users can be grouped according to various metrics including the quality of service (QoS) \cite{QOS31}, the level of received energy\cite{AngleDomain2019}, the maximum tolerable delay, etc. Most of the grouped transmissions fall under the category of the pull-based approach \cite{DN2015}, where every group has its unique group identification (GID) and the users of a group will access the BS if and only if their GID is granted by the paging message sent from the BS \cite{TsoukaneriGroup}. In \cite{Chih2011Energy}, every group further selects one of its members as the group head (GH). The GH will act as a relay node for other members in the same group.In general, grouped random access demonstrated some particular advantages.
However, these pull-based schemes may cause a serious latency problem while the interval between two paging messages that grant the same GID is quite long\cite{WGroup7}.

In this paper, we more focus on solving the active user detection challenge encountered in mMTC scenarios. \textcolor[rgb]{0,0,0}{More specifically, this paper aims at reducing the CS signaling overhead, saving the random access energy consumption, and accommodating significantly more cellular users}. Hence, a \textcolor[rgb]{0,0,0}{layered grouping RA scheme based on dynamic preamble selection} is proposed. The main contributions of this paper are summarized as follows:
\begin{enumerate}
	\item A layered grouping network framework is presented, where a cell is divided into several large clusters based on their geographical locations, and each cluster is further partitioned into a number of small groups according to the proposed construction and maintenance algorithms. It is assumed that users are \textcolor[rgb]{0,0,0}{normally} active \textcolor[rgb]{0,0,0}{in units of small groups,} not individually.

	\item In \textcolor[rgb]{0,0,0}{this layered grouping cellular architecture}, the initialization, update, as well as group head (GH) selection procedure of every small group are implemented autonomously, where a self-organizing \textcolor[rgb]{0,0,0}{optimum energy consumption (Opt-EC)} based K-means algorithm is designed and employed. The user which is capable of maximizing the energy efficiency of the entire group is selected as the GH.

	\item \textcolor[rgb]{0,0,0}{Associated with the layered cellular architecture}, the RA procedure is divided into cluster load estimation phase \textcolor[rgb]{0,0,0}{(namely, phase-I, in which RA operates in a push-based manner)} and active group detection phase \textcolor[rgb]{0,0,0}{(namely, phase-II, in which RA operates in a pull-based manner). The} conventional user ID (UID) based random access is replaced by a unique group ID (GID) based random access. \textcolor[rgb]{0,0,0}{Here, the layered cellular architecture, the two-phase access procedure, as well as the formed groups access entity, allow a BS to connect with much more coexisted users in an energy efficient manner.}

	\item Two kinds of preambles, short and long preambles, are employed, where the short preambles are orthogonal and \textcolor[rgb]{0,0,0}{allocated to clusters as their signature, while the long preambles having the cluster-load depended minimum length are non-orthogonal and allocated to group heads as group identity.} The state-of-the-art approximated message passing (AMP) algorithm \cite{Senel2018Grant, PartI2018} is employed for realizing the active group detection.
	
	\item To analyze the overhead problem and the impact of the preamble length in the AMP algorithm, a tight lower bound on the minimum preamble length (MPL) is derived based on the state-evolution method.

	\item \textcolor[rgb]{0,0,0}{Based on the preamble categorization and MPL lower bound analysis, a dynamic preamble selection (DPS) strategy is adopted in phase-II, where the required preambles having the cluster-load depended minimum length  are dynamically selected.} It is shown that, benefiting from both the hierarchical preamble assignment and \textcolor[rgb]{0,0,0}{the DPS} strategy, the \textcolor[rgb]{0,0,0}{overhead} of the proposed \textcolor[rgb]{0,0,0}{RA} strategy can be significantly \textcolor[rgb]{0,0,0}{reduced}.
\end{enumerate}

The rest of this paper is organized as follows: in Section \ref{sec:System Model}, the layered grouping network framework is proposed. \textcolor[rgb]{0,0,0}{Its} associated constructing and maintaining mechanism is designed, while the two-phase RA scheme is also depicted. Then, critical techniques employed in the proposed \textcolor[rgb]{0,0,0}{two-phase DPS RA strategy} including optimal energy computing (Opt-EC) based K-means grouping algorithm, AMP detection algorithm, and dynamic preamble selection algorithm are discussed in Section \ref{sec:Key}. The tight lower bound on the minimum preamble length is derived in Section \ref{The}. Simulation results are demonstrated and analyzed in Section \ref{sec:simulation}. Finally, the paper is concluded in Section \ref{sec:conclu}.

\section{System Model}
\label{sec:System Model}
\subsection{Network Framework of Layered Grouping }
\label{preaccess}

We consider an mMTC cell having a radius $R$, where the \textcolor[rgb]{0,0,0}{BS} equipped with a single antenna locates in the cell center and a total number of $N$ coexisted users randomly distribute in the cell. It is assumed that all the mMTC users are mainly static, a typical scenario in mMTC applications \cite{Sharma2020Toward}, e.g. the interactions among machines in the industrial automation, the monitoring in smart agriculture, the environment monitoring for public safety, etc.

\begin{figure}
	\begin{center}
		\includegraphics[width=0.9\textwidth]{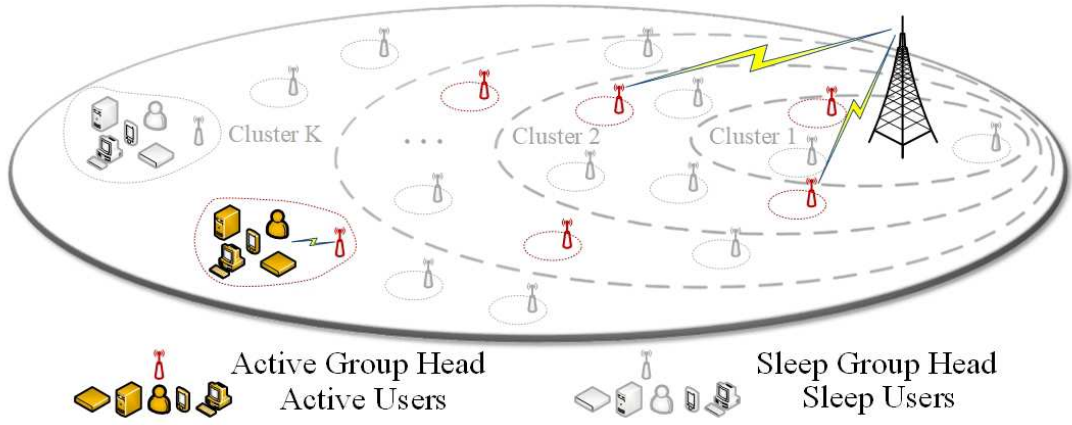}
		\caption{Network topology of the layered grouping in an mMTC cell.}
		\label{fig1SystemModle}
		\vspace{-0.1cm}
	\end{center}
\end{figure}

According to predefined system configurations including cell size, QoS requirement, maximum number of coexisted cellular users, etc. , BS will divide all the cellular users into a number of $K$ clusters. \textcolor[rgb]{0,0,0}{Typically, $K$ could be a small value, e.g. $K = 2,4,8,$ or $16$.} During phase-I of the proposed RA, in order to reduce the overhead, if the number of clusters is relatively small, then every cluster can be distinguished by a very short orthogonal preamble. Hence, the overhead is reduced. On the other hand, in order to facilitate the practical synchronization, it is better that the users in the same cluster experience similar transmission delays, normally located in a geographical area having roughly the same distance to the BS. This could be achieved by estimating the received signal strength (RSS) at the BS \cite{Lau07Enhanced}, \cite{Bhattacharjee19Weather}. Ideally, the entire cellular is divided into a number of $K$ rings and all the users located in the same ring will be assigned to the same cluster. This effect is visualized by the dashed ellipses in Fig.\ref{fig1SystemModle}. In practice, owing to the limited localization accuracy, some edge users may be assigned to their adjacent cluster. But this potential mismatch will not impact the proposed two-phase GF RA scheme, as the proposed RA scheme is capable of adapting to unbalanced cluster loads.

Then, the users assigned to the $k^{th}$ cluster $c_k$, $k=1, 2, \cdots, K$ will further participate in a number of $M^{(k)}$ groups. These groups are initialized and updated in a self-organizing manner. The $m^{th}$ group in $k^{th}$ cluster is denoted by $g_{k,m}$. The users pertaining to a group is termed as group members and the set of their user ID is denoted by $\mathcal{G}_{k,m}$. The number of group members is termed as the group size and denoted by $|g_{k.m}|$. We would like to constrain the group size by a small value. Because a large group size normally results in longer distances between GH and its group members, which implies less reliable \textcolor[rgb]{0,0,0}{device-to-device (D2D)} links. Actually, it was shown in Fig.3 of \cite{8847234} that packet error rate over D2D links becomes non-negligible after the group size exceeds 20.
Hence we could further reasonably assume that all the group members are close to each other. As a benefit, the internal message exchange among group members can be reliably realized by the \textcolor[rgb]{0,0,0}{D2D} communication technique \textcolor[rgb]{0,0,0}{\cite{8847234}}, which could be interference-free to other groups and hence spectral efficient. Moreover, a particular group member, namely $\dot{u}_{k,m}$ will be selected as the GH of  group $g_{k,m}$. If a normal group member $u_{n}$ wants to communicate with the BS, it firstly sends the message to its GH $\dot u_{k,m}$. Then the GH $\dot u_{k,m}$ relays this message to BS, and vice versa. It implies that throughout the GF RA procedure, the GH $\dot u_{k,m}$  will communicate with the BS on behalf of all the members in the same group. The above-mentioned mechanism is visualized by the dotted ellipses in Fig.\ref{fig1SystemModle}, where the GH is equivalently denoted as a relay node.

Finally, a GH $\dot u_{k,m}$ possesses two kinds of access preambles. The first one, namely $\mathbf{s}_k^I$ is only used in phase-I and actually the signature of cluster $c_k$. All cluster signatures are orthogonal to each other, i.e.
$\langle{\mathbf{s}_i^I, \mathbf{s}_j^I\rangle}=\left\{
\begin{aligned}
0,\quad if\ j\neq i\\
1,\quad if\ j=i
\end{aligned}
\right.
$
, where $i,j=1, 2, \cdots, K$. Since $K$ is relatively small, this orthogonality can be easily satisfied, even for a short preamble length. The second one, namely $\mathbf{s}_{k,m}^{II}$ is only used in phase-II and actually the signature of group $g_{k,m}$. Since $M^{(k)}$ is normally a very large number in mMTC scenarios, $\mathbf{s}_{k,m}^{II}$ employed by different groups in the same cluster has to be non-orthogonal sequences for reducing overhead. More details of preamble assignment could be found in Section \ref{section2B} and Section \ref{dpa}.

\subsection{Construction and Maintenance of the Layered Grouping }
\label{section2B}
The construction of clusters can be controlled by BS, where only an approximated distance from a user to BS is required. On the other hand, the formation and update of groups in each cluster may also be implemented in a centralized manner \cite{8847234} \cite{7992937}, where BS controls the selection of GHs and assigns their group members. However, this centralized management requires a range of global information including users' accurate positions, propagations, data rates, battery levels, etc. Aggregating these information from millions devices in the mMTC scenario may become prohibitive. Hence distributed self-organized formation and update of groups are advocated in this paper. Thus, the construction and maintenance procedures of the proposed \textcolor[rgb]{0,0,0}{layered} network framework are designed as follows
\begin{enumerate}
\item While a user (device) $u_{n}, n = 1,2,\ldots, N,$ firstly powers on in the cell and hears the system broadcast information (including the power level of control channel) from BS, its registration process will then start by sending a registration message containing user ID, device type, and a couple of reference signals, to BS in a contention free manner\footnote{Since the number of users simultaneously switching on is normally extremely low, contention-free transmission of registration message could be realized by predefining a small set of specific channel resources.}.
\item Based on the reference signals contained in the registration message, BS is capable of approximating the distance between a user and itself by utilizing RSS aided positioning techniques \cite{Lau07Enhanced}, \cite{Bhattacharjee19Weather} and further assigning $u_{n} $ to an appropriate cluster $c_{k}$. Then, BS assigns the generation method of a pair of preambles $\mathbf{s}^{I}_{k} $  and $\mathbf{s}^{II}_{k,m} $ , the initial preamble lengths, as well as the group size $\left\vert g_{k,m} \right\vert $  to  $u_{n}$. These information and a couple of reference signals are contained in the registration response message (RRM).
\item Based on the reference signals in RRM, user  $u_{n}$ is capable of estimating the channel from BS, and the associated channel state information (CSI) is denoted by $h_{n,b} $. It is assumed that all the channels are reciprocal and have a relative long coherent time based on the fact that the mMTC devices are mainly static in our application scenarios. According to RRM, user $u_{n}$ becomes aware of its cluster index $k$. Then, user $u_{n}$  will further autonomously select itself as a GH in a probability of $\frac{1}{\left\vert g_{k,m} \right\vert} $ .
\item BS will periodically broadcast a group update opportunity message (GUOM) to all cellular users. Bearing the quasi-static property of our application scenario in mind, the group update period could be generally long, say daily or even weekly for reducing system overhead.
\item Once the cellular users hear GUOM, they will implement the group initialization\footnote{In the case a group has not been created before.} or update\footnote{In the case a group has existed.} procedure in a self-organizing manner and via D2D links. An Opt-EC based K-means grouping algorithm is designed to iteratively improve the grouping relationship and select the energy efficient GHs, which will be elaborated in Section \ref{OPTECKMEANS}. With the aid of this K-means grouping algorithm, the groups in a cluster are constructed and updated.
\item If the role of a user changes (i.e. switches from a normal group member to a GH and vice versa), it will inform BS of its new state. The BS will add or remove the associated group ID from its group list.
\end{enumerate}

\begin{figure}[htbp]
	\begin{center}
		\includegraphics[width=0.5\textwidth]{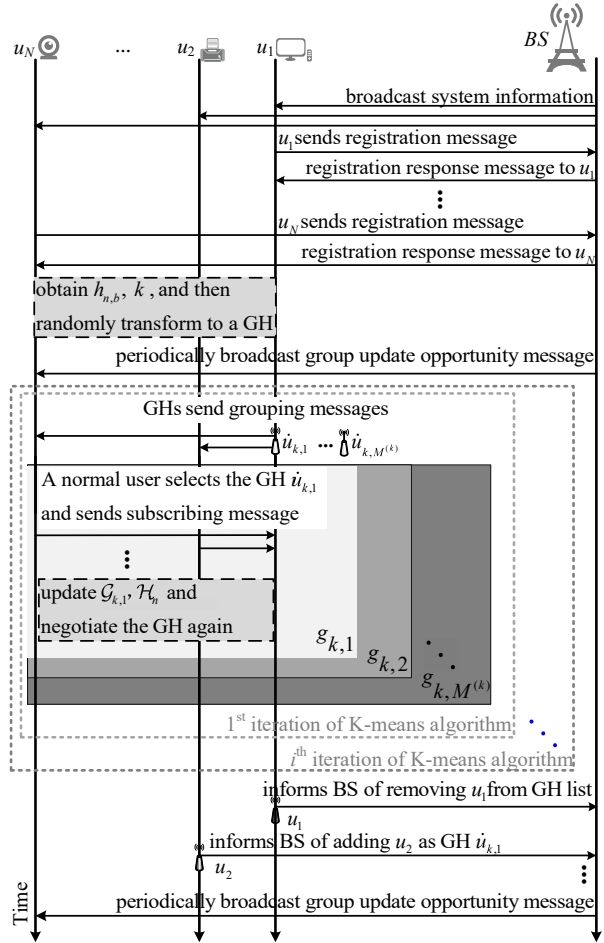}
		\caption{Construction and maintenance of the layered grouping network framework.}
		\label{fig2Networking}
		\vspace{-0.1cm}
	\end{center}
\end{figure}

The above-mentioned construction and maintenance procedures are illustrated in Fig.\ref{fig2Networking}.

\subsection{Two-Phase Random Access}
\label{RA}

\begin{figure}[htbp]
	\begin{center}
		\includegraphics[width=0.5\textwidth]{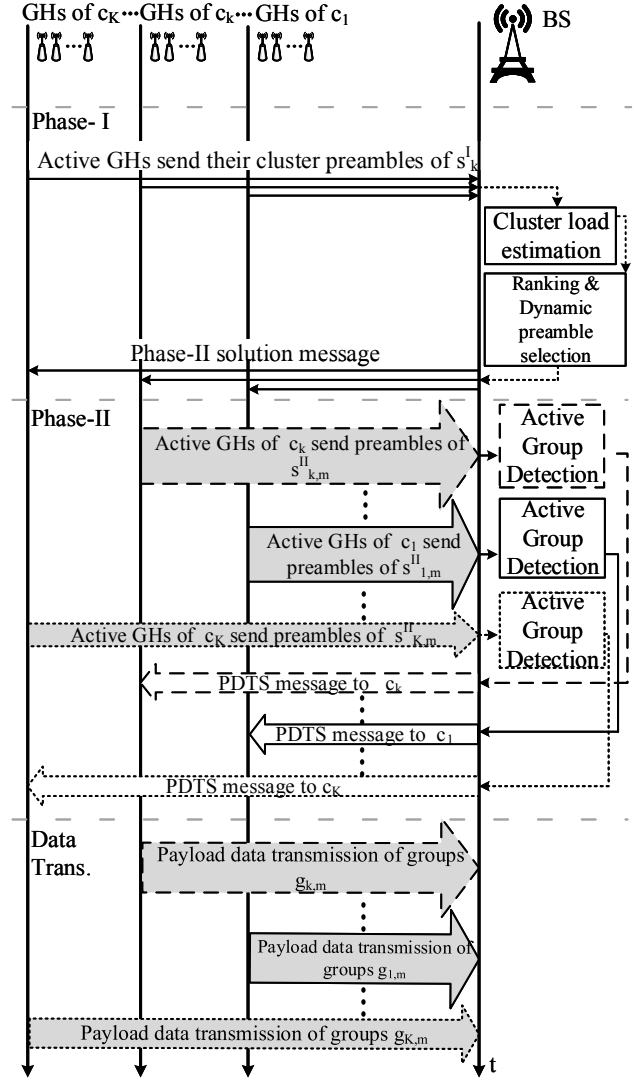}
		\caption{Transmission stream of the two-phase grant-free random access procedure.}
		\label{fig3RandomAccess}
		\vspace{-0.1cm}
	\end{center}
\end{figure}

As depicted in Fig.\ref{fig3RandomAccess}, we divide the proposed RA procedure into phase-I and phase-II. During phase-I, after receiving the RAO message from the BS, the GHs of all the active groups in the cell will first transmit their cluster preambles $\mathbf{s}_k^{I}$. Bear in mind that $\mathbf{s}_k^{I}$ of a GH $\dot u_{k,m}$ has been specified during the registration process introduced in Section \ref{section2B}. Furthermore, a group $g_{k,m}$ is regarded as an active group if one or more members in this group want to transmit payload data to the BS. Accordingly, the signal received at the BS during phase-I can be written as
\begin{equation}
	\label{yc}
	\mathbf{y}^I=\sum_{k=1}^K \sum_{m=1}^{M^{(k)}} a_{k,m}\sqrt{P_{k,m}}\mathbf{s}_k^Ih_{k,m\rightarrow b}+\mathbf{\omega},
\end{equation}
where the activity state $a_{k,m}\in\{0,1\}$ indicates the activity of a  group $g_{k,m}$. If $g_{k,m}$ is active, $a_{k,m}=1$, and so forth. $h_{k,m\rightarrow b}$ is the CSI of the channel from the GH $\dot u_{k,m}$ to the BS, which simultaneously encapsulates both the large-scale fading and small-scale fading. $\mathbf{\omega}$ is an AWGN vector, whose elements obey i.i.d complex Gaussian distribution having zero mean and variance $\sigma^2$. $P_{k,m}$ is the actual transmit power of the GH $\dot u_{k,m}$, which is given by
\begin{equation}
	\label{rho}
	P_{k,m}=P\cdot\frac{\beta_{min}}{\beta_{k,m}},
\end{equation}
where $P$ is a common transmit power that can be afforded by all the GHs. The value of $P$ could be assigned to the users during their registration process. $\beta_{k,m}$ is the average large-scale fading coefficient of the channel from the GH $\dot u_{k,m}$ to the BS. It could be continuously updated by testing the reference signals transmitted by the BS, e.g. the reference signals included in the registration response message, the GUO message and the RAO message. $\beta_{min}$ is the minimum value among $\beta_{k,m}, k\in\{1, 2, \cdots, K\}, m\in\{1, 2, \cdots, M^{(k)}\}$, which could be carried in the RAO message. By substituting \eqref{rho} to \eqref{yc}, it is apparent that specifying the actual transmit power $P_{k,m}$ according to \eqref{rho} is equivalent to employing an adaptive power control mechanism. Hence, at the BS, the average power of the signal received from $\dot u_{k,m}$ approximately equals to $P\cdot \beta_{min}$ \cite{Senel2018Grant}.

During a RAO, the number of active groups in a cluster $c_k$ is termed as its cluster load. Hence, based on the received signal $\mathbf{y}^I$,  the BS is capable of estimating the cluster load of $c_k$, which is given by
\begin{equation}
\label{load}
	\hat{M}_{ac}^{(k)} = \frac{\langle{\mathbf{y}^I,\mathbf{s}_k^I}\rangle}{\sqrt{P\beta_{min}}}.
\end{equation}
 Correspondingly, the sparsity of cluster $c_k$ is approximated by
\begin{equation}
	\hat{\lambda}_k=\frac{\hat{M}_{ac}^{(k)}}{M^{(k)}}.
\end{equation}

After the cluster load estimation formulated by \eqref{load}, the BS could rank the access priority of different clusters according to their cluster loads $\hat{M}_{ac}^{(k)}$.  A higher access priority is assigned to the cluster having a larger cluster load. As indicated by the \textcolor[rgb]{0,0,0}{largest} streams in the middle and bottom of Fig.\ref{fig3RandomAccess}, the GHs in the highest loaded cluster will firstly send their group access preambles and their payload data, respectively. Based on the cluster load estimation, as well as the compressive sensing theorem, the BS will adaptively select the group preamble length of $L_{k} = |\mathbf{s}^{II}_{k,m} |$ for different clusters. After ranking the access priority and selecting the group preamble length, the BS could arrange the access slots of every cluster. Then, the cluster-specific access slots and preamble lengths that will be used in phase-II are broadcasted by the BS. This message is called as the ``Phase-II solution message'' (PSM) in Fig.\ref{fig3RandomAccess}. More details of \textcolor[rgb]{0,0,0}{DPS strategy} is provided in Section \ref{dpa}.

After receiving the phase-II solution messages, the GHs of all the active groups will generate their unique preambles of $\mathbf{s}_{k,m}^{II}$. The associated generation method has been determined in the user registration process\footnote{In this paper, complex gaussian sequences having zero mean and variance $\frac{1}{\left|\mathbf{s}_{k,m}^{II}\right|}$ are employed as $\mathbf{s}_{k,m}^{II}$.} and the preamble length is indicated by the phase-II solution messages. At this moment, the proposed \textcolor[rgb]{0,0,0}{ two-phase} RA procedure starts its phase-II operations. Firstly, the GHs of all the active groups in the same cluster will simultaneously send their group preambles during a specific access slot that has been indicated by the phase-II solution messages, which are illustrated by the shadowed arrows having  dashed, solid, and dotted borders in the middle of Fig.\ref{fig3RandomAccess}. During the access slot of cluster $c_k$, the signal received at the BS is given by
\begin{equation}
	\label{yg}
	\mathbf{y}_k^{II}=\sum_{m=1}^{M^{(k)}}a_{k,m}\sqrt{P_{k}}\mathbf{s}_{k,m}^{II}h_{k,m\rightarrow b}+\mathbf{\omega}_k =\sqrt{P_{k}}\mathbf{S}_{k}^{II}\mathbf{x}_{k}+\mathbf{\omega}_k,
\end{equation}
where $a_{k,m}$ and $h_{k,m \rightarrow b}$ have been defined in \eqref{yc}. $\mathbf{\omega}_k$ is the AWGN vector in the access slot of $c_k$. $P_{k}$ is the standard transmission power of all the GHs in phase-II. Let $L_k$ denote the length of $\mathbf{s}_{k,m}^{II}$, then we have $\mathbf{S}_{k}^{II}=[\mathbf{s}_{k,1}^{II},\mathbf{s}_{k,2}^{II},\cdots,\mathbf{s}_{k,M^{(k)}}^{II}]\in\mathbb{R}^{L_k\times M^{(k)}}$ and $\mathbf{x}_{k}=[x_{k,1},x_{k,2},\cdots,x_{k,M^{(k)}}]^T$, where $x_{k,m}=a_{k,m}h_{k,m\rightarrow b}$.

As illustrated in the middle of Fig.\ref{fig3RandomAccess}, after the active group detection of all the clusters are completed, the BS will broadcast payload data transmission solution (PDTS) message to every cluster. The active group ID detected by the BS and its payload data transmission time slot are carried in the PDTS message. Finally, as observed at the bottom of Fig.\ref{fig3RandomAccess} that the GH of every active group will relay the payload data of their group members to the BS at different time slots as indicated by PDTS messages.

\section{Group Maintenance and Detection Algorithms}
\label{sec:Key}
\subsection{Group Initialization and Update Algorithms}
\label{OPTECKMEANS}
As stated in Section \ref{section2B}, during step 5 of the construction and maintenance procedure of the layered grouping framework, after receiving the group update opportunity message from BS, the existing GHs $\dot{u}_{k,m}, 1\leq m \leq M^{(k)}$ will broadcast grouping messages. Group ID is contained in these grouping messages. A registered neighboring non-GH user $u_{n}$ will select the one from which it hears the strongest grouping message as its intended GH. Then, $u_{n}$ \textcolor[rgb]{0,0,0}{broadcasts} a subscribing message (SM), which carries its user ID, intended group ID. After receiving all the subscribing messages, a GH $\dot{u}_{k,m}$ is capable of determining all of its group members, and then feedbacks this temporary decision of $\mathcal{G}_{k,m}$ to all the group members. The transmission of above-mentioned intra-group signaling messages will rely on dedicated spectrum such as 5905-5925 MHz in 5G NR V2X (PC5 interface) \cite{V2Xstandard}, or unlicensed spectrum technologies, namely D2D outband communications \cite{8847234}. Hence, the impact of these extra overheads on the cellular radio access network (RAN) can be negligible. These intra-group signaling are summarized as lines 3-5 in Algorithm \ref{algK}.

The proposed Opt-EC based K-means grouping algorithm
aims at selecting the best GH, which simultaneously minimizes the energy required by intra-group communications\footnote{Because, during the ensuing payload transmission slot, the group members will first send their data to the GH, then the GH relays all the data to the BS.} and that required by external \textcolor[rgb]{0,0,0}{cellular communications.} In order to realize this optimization, every group member should be aware of all the channel conditions from other group members to itself, namely $\mathcal{H}_n=\{h_{i,n}\ :\ n\in \mathcal{G}_{k,m},\ i=1, 2, \cdots, \left|g_{k,m}\right|,\ i\neq n\}$.
Again, this requirement could be satisfied by exploiting D2D outband communications and the associated overhead is negligible for cellular RAN. The update of $\mathcal{H}_n$ for every group member is summarized as line 6 in  Algorithm \ref{algK}.

\begin{algorithm}[t]
	\caption{Opt-EC based K-means grouping algorithm}	
	\label{algK}
	\SetKwInput{KwOut}{Initialize}	
	\KwOut{cluster index $k$.}
	\Begin{
	\nlset{1}\While{ (Predefined \textcolor[rgb]{0,0,0}{max} iterations has not been reached.)}
    {		\lnlset{loopbegin}{2}
            \For{($m=1$ \KwTo $M^{(k)}$)}{
				 \nlset{3}$\dot{u}_{k,m}$ selected in last iteration broadcasts grouping message;
			}
            \For{($n=1$ to $N$)}{
			\For{($m=1$ \KwTo $M^{(k)}$)}{
				 \nlset{4}$u_n$ attempts to hear from $\dot u_{k,m}$ and estimate $h_{\dot u_{k,m},n}$;
			}
			 \nlset{5}$ u_n \xrightarrow{subscribe \ group} \mathop{\arg\max}\limits_{\dot u_{k,m}\ heard\ by\ u_n}h_{\dot u_{k,m},n}$;
			}
		
            \For{$m=1$ to $M^{(k)}$}{
  		     \nlset{6} Update $\mathcal{G}_{k,m}$ and every $\mathcal{H}_n$; \\
			 \lnlset{loopend}{7}$\dot u_{k,m} = \mathop{\arg\min}\limits_{u_n \in \mathcal{G}_{k,m}} \gamma_{k,m}(u_n)$;
		}
	}}
\end{algorithm}

Herein, we further assume that the transmit power, packet size and bandwidth of the reference signals contained in the intra-group signaling messages are fixed to $P$, $D$, and $B$, respectively.
Assume that $u_n$, $u_i$ are group members of $g_{k,m}$, i.e. $n,i\in\mathcal{G}_{k,m}$. The achievable error-free data-rate from $u_i$ to $u_n$ could be characterized by \textcolor[rgb]{0,0,0}{$R_{i,n}=B\log_2\left[1+\frac{P\left|h_{i,n}\right|^2}{N_o}\right]$, where $N_o$ denotes the power density of additive Gaussian noise. Accordingly, if we select $u_n$ as the GH $\dot u_{k,m}$, the energy required by intra-group communications in ensuing payload transmission slot could be characterized by $\epsilon^{(n)}_{\rm{inner}} = \sum_{i\in \mathcal{G}_{k,m}\atop i\neq n} P\cdot \frac{D}{R_{i,n}}.$}

\textcolor[rgb]{0,0,0}{Similarly, the achievable error-free data-rate from $u_n$ to BS could be characterized by $R_{n,b}=B\log_2\left[1+\frac{P\left|h_{n,b}\right|^2}{N_o}\right]$. Again, if we select $u_n$  as GH $\dot u_{k,m}$, the energy required by external communications between $u_n$ and BS could be characterized by $\epsilon^{(n)}_{\rm{outer}} = P\cdot \frac{\vert g_{k,m} \vert \cdot D}{R_{n,b}}.$}

\textcolor[rgb]{0,0,0}{Finally, the energy efficiency of selecting $u_n$ as the GH $\dot u_{k,m}$ could be characterized by $\gamma_{k,m}(u_n)=\epsilon^{(n)}_{\rm{inner}}+\epsilon^{(n)}_{\rm{outer}}$, where a smaller value of $\gamma_{k,m}(u_n)$ implies a better energy efficiency. $\gamma_{k,m}(u_n),\ n\in\mathcal{G}_{k,m}$ will be calculated at the group member $u_n$ and then be forwarded to current GH. Hence, by running the Opt-EC based K-means algorithm, the GH of $g_{k,m}$ could be selected according to $\dot u_{k,m}=\mathop{\arg\min}\limits_{u_n\in\mathcal{G}_{k,m}}\gamma_{k,m}(u_n),$ which is summarized as line 7 in Algorithm \ref{algK}.}

The above-mentioned operations can be repeated again among the cellular users for further adjusting the grouping relationships and optimizing GH selections. But, in practice, owing to the limited energy budget, the iterations of Algorithm \ref{algK} has to be terminated within a predefined maximum number.

\subsection{MMSE Denoiser Based AMP Algorithm}
\label{amp}

The classical system model used in compressive sensing is represented as
\begin{equation} \label{ampmodel}
	\mathbf{y}=\mathbf{Ax+\omega},
\end{equation}
where $\mathbf{x}$ is the original signal vector having a number of $M$ elements. $\mathbf{A}\in \mathbb{R}^{L\times M}$ is the measurement matrix. $\omega$ is an AWGN vector. \textcolor[rgb]{0,0,0}{Since $M \gg  L$}, $\mathbf{y}$ is actually a compressed and corrupted observation of $\mathbf{x}$. As an efficient solution of \textcolor[rgb]{0,0,0}{recovering $\mathbf{x}$ from $\mathbf{y}$}, the approximated message passing (AMP) algorithm is first proposed in \cite{AMP2009}. Its theoretical derivation could be found in \cite{DonohoHow}.

In more detail, the AMP algorithm could be formulated by the following iterative procedures
\begin{equation}\label{eq4}
\mathbf{x}^{t+1} = \eta_t(\mathbf{A}^*\mathbf{z}^t+\mathbf{x}^t,\mathbf{\tau}_t),
\end{equation}
\vspace{-0.3cm}
\begin{equation}\label{eq5}
\mathbf{z}^{t+1} = \mathbf{y}-\mathbf{A}\mathbf{x}^{t+1}+\frac1 \mu \mathbf{z}^{t}\langle \eta^{'}_t(\mathbf{A}^*\mathbf{z}^{t}+\mathbf{x}^{t};\mathbf{\tau}_{t})\rangle,
\end{equation}
\vspace{-0.1cm}
\begin{equation}\label{eq6}
{\tau}_t \approx \frac1 {\sqrt L}\lVert \mathbf{z}^t \rVert _2.
\end{equation}
where $\mathbf{x}$ is initialized to a zero vector, i.e. $\mathbf{x}^0=0$, $\eta_t(\cdot)$ is the soft thresholding function and $t$ \textcolor[rgb]{0,0,0}{is} the index of iteration, $\mathbf{x}^t$ represents the estimation of $\mathbf{x}$ at the $t^{th}$ iteration, $\mathbf{z}^t$ calculates the residual component, $\mathbf{A}^*$ denotes the conjugate transpose of $\mathbf{A}$, $\langle\cdot\rangle$ denotes the average of a vector, $\eta^{'}_t$ is the first derivative of $\eta_t$ with respect to the first argument, and $\mu = \frac{L}{M}$ is the under-sampling ratio. In contrast to the conventional iterative thresholding algorithms, $\eta^{'}_t(\mathbf{A}^*\mathbf{z}^{t}+\mathbf{x}^{t};\mathbf{\tau}^{t})$ is a new component invoked by the AMP algorithm and known as the \textcolor[rgb]{0,0,0}{``Onsager reaction term''}, which is identified as the fundamental improvement of the AMP algorithm.

Furthermore, in \cite{Chen2018Sparse}, the soft thresholding denoiser $\eta_{t}\left( \cdot \right)$ is developed to an MMSE denoiser \textcolor[rgb]{0,0,0}{as follows
\begin{equation}\label{eq7}
\eta_t(\hat{x}_m^t,g_m) = \mathbb{E}[X|\hat{X}^t=\hat{x}_m^t,G=g_m],
\end{equation}
where $X$, \textcolor[rgb]{0,0,0}{$\hat{X}^t$, $\hat{x}^{t}_{m}$, and $g_{m}$ have the same definitions as that in} \cite{Chen2018Sparse}.} 

\textcolor[rgb]{0,0,0}{This} MMSE denoiser can employ the large-scale fading coefficient $G$ known at the BS as a priori information of AMP algorithm. Hence it results in a better recovery accuracy. The above-stated MMSE based AMP algorithm is employed to solve the active group detection problem by replacing the classical compressive sensing model given in \eqref{ampmodel} with the group access model given in \eqref{yg}. \textcolor[rgb]{0,0,0}{Accordingly,} the variables $\mathbf{y}$, $\mathbf{A}$, $\mathbf{x}$, $\mathbf{\omega}$ involved in \eqref{ampmodel}$\sim$\eqref{eq6} \textcolor[rgb]{0,0,0}{are} replaced by the variables $\mathbf{y}_k^{II}$, $\mathbf{S}_k^{II}$, $\mathbf{x}_k$, $\mathbf{\omega}_k$ given in \eqref{yg}, respectively. The under-sampling ratio of $\mu$ in \eqref{eq5} \textcolor[rgb]{0,0,0}{is} calculated by  $\frac{L_k}{M^{(k)}}$. The number of total elements $M$ and nonzero elements $M_{ac}$ in $\mathbf{x}$ is replaced by that of total groups $M^{(k)}$ and  active groups $M^{(k)}_{ac}$ in a cluster $c_{k}$, respectively.

\begin{algorithm}[t]
	\caption{Dynamic preamble selection algorithm.}
	\label{algD}
	\KwIn{$\bm{y}^I$}
	\KwOut{$\bm{V}$, $\bm{L}=[L_1, L_2, \cdots, L_K]$	\tcc*[r]{$\bm{V}$ indicates the access slot indices of every cluster, $\bm{L}$ indicates the preamble lengths selected for every cluster.}
	}
	\SetKwInput{KwOut}{Initialize}	
	\KwOut{$K$, target $pF$, target $pM$, $\left\lbrace M^{(k)}, k=1, 2, \cdots, K\right\rbrace$,
		 $\left\lbrace[R_1^{(k)}, R_2^{(k)}], k=1, 2, \cdots, K\right\rbrace$
		 	\tcc*[r]{$R_1^{(k)}, R_2^{(k)}$ are the inner and outer radius of the $k^{th}$ cluster, respectively.}}
	\BlankLine	
	\Begin{
	\nlset{1}\For{k=1 \KwTo $K$}{
		\nlset{2}$\hat{M}_{ac}^{(k)}=\mathrm{Estimate\_Cluster\_Load}(\bm{y}^I, k)$	\tcc*[r]{according to \eqref{load}.}		
		
		\nlset{3}$L_k=\mathrm{Lower\_Bound\_on\_MPL}(\hat{M}_{ac}^{(k)}, M^{(k)}, R_1^{(k)}, R_2^{(k)}, pF, pM)$;
		\tcc*[f]{according to \eqref{neweq5}, \eqref{neweq9} and \eqref{Lk3}.}	
			
		\nlset{4}$\bm{L}[k]=1.1*L_k$ \tcc*[r]{slightly enlarge $L_k$, see Section \ref{The}.}
		
		\nlset{5}$\hat{\mathbf{M}}_{ac}[k] = \hat{M}_{ac}^{(k)}$;
	}
	\nlset{6}$\bm{V}=\mathrm{Allocate\_Access\_Slot}(\hat{\mathbf{M}}_{ac}, \bm{L})$
	\tcc*[r]{arrange the access priority of every cluster in the descending order of $\hat{M}_{ac}^{(k)}$, then allocate the access slot indices of every cluster according to $\bm{L}$.}
	
	\nlset{7}\KwSty{Return:} $\bm{V}$, $\bm{L}$;		
}
\end{algorithm}

\subsection{Dynamic Preamble Selection Algorithm}
\label{dpa}
In the context of our active group \textcolor[rgb]{0,0,0}{detection and according to the CS theorem \cite{DecodingTao2005,Eldar2012Compressed},} to satisfy a certain detection accuracy, the minimum preamble length is related to the number of active groups in a cluster, namely $M_{ac}^{(k)}$ and to the total number of groups in a cluster, namely $M^{(k)}$. Apparently, in practice, the value of $M_{ac}^{(k)}$ and $M^{(k)}$ shall be salient different in different clusters. It implies employing a preamble with inappropriate length will result in either a serious detection inaccuracy or an excessive overhead. Accordingly, a \textcolor[rgb]{0,0,0}{DPS} algorithm shown in algorithm \ref{algD} is designed and employed in phase-II of the random access. \textcolor[rgb]{0,0,0}{The technical challenge of algorithm \ref{algD} occurs at its line 3 that evaluates the MPL. Owing to analyzing complexity and importance, we specifically discuss it in Section \ref{The}.} 

\section{Theoretical Analysis on the Minimum Preamble Length}
\label{The}

\begin{figure}[htbp]
	\begin{center}
		\includegraphics[width=0.5\textwidth]{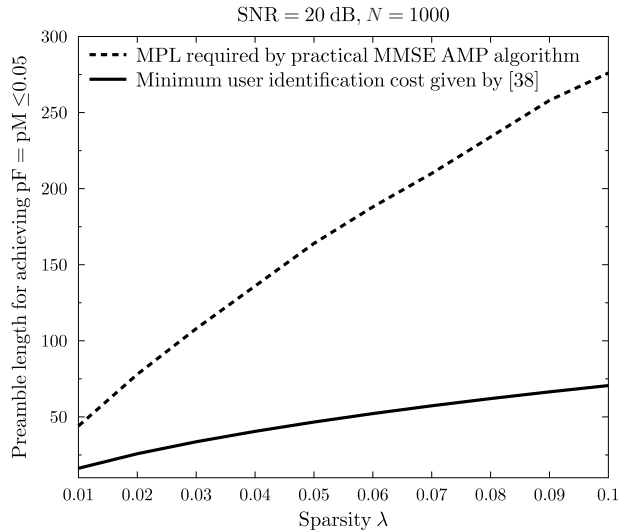}
		\caption{MPL versus sparsity. The MPL required by \textcolor[rgb]{0,0,0}{MMSE-AMP} algorithm and that predicted by \cite{Chen17Capacity} are compared. AWGN channels are assumed.}
		\label{fig4LBCompares}
		\vspace{-0.1cm}
	\end{center}
\end{figure}

According to Section \ref{dpa}, finding the MPL required by MMSE based AMP algorithm for achieving a certain data recovery accuracy makes great sense. Similar works have been attempted in \cite{DecodingTao2005} and \cite{Eldar2012Compressed}. However, two deficiencies of the MPL calculation method given in \cite{Eldar2012Compressed} prevent us from applying it in \textcolor[rgb]{0,0,0}{our DPS algorithm:} (1) two constant parameters, namely $C_1$ and $C_2$ are involved, i.e., instead of an exact value, it only provides an asymptotical order; (2) it does not relate to a specific data reconstruction method.
\textcolor[rgb]{0,0,0}{Furthermore}, the authors of \cite{Aksoylar13Sparse} and \cite{Chen17Capacity} attempted to answer this fundamental question from the perspective of classical asymptotic information theoretic analysis. In \cite{Chen17Capacity}, seeking for the MPL is termed as the “minimum user-identification cost” problem. In their Gaussian many-access channel (MnAC) model, the minimum number of channel uses for guaranteeing an error-free random user identification is given by
\begin{equation}\label{equx}
	L=\frac{N\cdot H_2(\frac{N_{ac}}{N})}{\frac{1}{2}\log(1+N_{ac}\gamma)},
\end{equation}
where $N$ denotes the total number of cellular users. In contrast, $N_{ac}$ denotes the average number of active cellular users. $\gamma$ denotes the signal-to-noise ratio (SNR) and it is assumed in \cite{Chen17Capacity} that every active user is subject to the same power constraint of $\gamma$. Besides, the entropy function is defined as $H_2(\mathnormal{p})=-\mathnormal{p}\log(\mathnormal{p})-(1-\mathnormal{p})\log(1-\mathnormal{p})$. However, the theorems provided in \cite{Chen17Capacity} are \textcolor[rgb]{0,0,0}{still not} suitable in our scenarios owing to two reasons: (1) the result shown in \eqref{equx} does not relate to any specific active user detection algorithm, hence a significant gap may exist between the MPL required by AMP algorithm and that predicted by \eqref{equx}, as illustrated in Fig.\ref{fig4LBCompares}; (2) only Gaussian noisy channels are considered. However, both large-scale and small-scale fading effects are taken into account in our system for modelling a more practical random access scenario.

In the following, we will provide a tight lower bound on MPL \textcolor[rgb]{0,0,0}{for MMSE-AMP algorithm.}
The state evolution method proposed in \cite{Chen2018Sparse} is employed, where the mean square error (MSE) of data reconstruction is regarded as a state variable. In more detail, at every iteration of MMSE based AMP algorithm, $\hat{X}^t$ in \eqref{eq7} is modeled as a noise corrupted signal. Hence, $\hat{X}^t$ could be formulated as
\begin{equation}\label{neweq0}
\hat{X}^t = X +\tau_t\cdot \omega_0,
\end{equation}
where the random variable $\omega_0$ follows complex Gaussian distribution with zero mean and unit variance. Then, $\tau_t$ is \textcolor[rgb]{0,0,0}{referred to} the variance of the $t^{th}$ estimation $\hat{X}^t$. Particularly, according to \cite{Chen2018Sparse}, $\tau_t$ is given by
\begin{equation}\label{taut2}
\tau_{t+1}^2 = \frac {\sigma^2} {P_kL_k}+ \frac {M^{(k)}} {L_k} MSE(\tau_t),
\end{equation}
where $\sigma^2$ is the variance of background noise $\omega$ involved in \eqref{yc}. The function $MSE(\cdot)$ evaluates the \textcolor[rgb]{0,0,0}{MSE} of its input variable and is specified in \cite{Chen2018Sparse}.

Based on the state evolution method \cite{Chen2018Sparse}, in order to achieve a high data reconstruction accuracy, the recursive reconstruction progress formulated by \eqref{eq4}-\eqref{eq6} has to converge. It means the variance of the AMP estimation, namely $\tau_t$ should constantly decrease to $\frac {\sigma^2} {P_kL_k}$. Hence, the following inequality holds
\begin{equation}\label{taut21}
\tau_{t+1}^2 \leq \tau_t^2, \forall t.
\end{equation}
By substituting \eqref{taut2} into the inequality \eqref{taut21}, a lower bound on preamble length for satisfying the convergence of AMP algorithm is given by
\begin{equation}\label{Lk}
L_k \geq \frac {\frac {\sigma^2} {P_k}+ {M^{(k)}}MSE(\tau_t)} {\tau_t^2}.
\end{equation}

In order to evaluate the performance of AMP algorithm, two metrics are invoked: (1) the probability of missed detection (pM) in cluster $c_k$; (2) the probability of false alarm (pF) in cluster $c_k$. They are defined as
\begin{equation}\label{eqpM}
pM^{(k)} = \frac{\sum_{m=1}^{M^{(k)}}\{\hat{x}_{k,m}< \theta \ \&\ x_{k,m}\neq 0\}}{\sum_{m=1}^{M^{(k)}}\{x_{k,m}\neq 0\}},
\end{equation}
\begin{equation}\label{eqpF}
pF^{(k)} = \frac{\sum_{m=1}^{M^{(k)}}\{\hat{x}_{k,m} \geq \theta\ \&\ x_{k,m}= 0\}}{\sum_{m=1}^{M^{(k)}}\{x_{k,m}= 0\}},
\end{equation}
where $x_{k,m}$ is defined in \eqref{yg}, $\hat{x}_{k,m}$ is the estimation of $x_{k,m}$ given by the AMP based active group detection, and $\theta$ denotes the decision threshold employed by AMP algorithm. While $\hat{x}_{k,m} \geq \theta $, AMP algorithm will regard $g_{k,m}$ as an active group. Then, according to the state evolution method, the $\mathrm{pM}^{\left( k \right)}$ and $\mathrm{pF}^{\left( k \right)}$ \textcolor[rgb]{0,0,0}{that can} be achieved in the $t^{\mathrm{th}}$ AMP iteration could be characterized by
\begin{equation}\label{neweq5}
\left\{
    \begin{array}{l}
           pF^{(k)} = e^{-\frac{\theta^2}{\tau_t^2}}, \\
           pM^{(k)} =  \frac 1 M \sum \limits_{m=1}^M(1-e^{-\frac{\theta^2}{\tau_t^2+g_{m}^2}}) = \displaystyle{\int (1-e^{-\frac{\theta^2}{\tau_t^2+g^2}})\cdot P_G^{(k)}(g)\ dg} ,
        \end{array}
\right.
\end{equation}
where $P_G^{(k)}(g)$ is the probability density function of the large-scale fading coefficient $g$. The random variable $g$ takes both the path-loss effect and the shadowing effect into account. In more detail, the path-loss effect is modeled as $\alpha+\beta log_{10}(d)$, where $d$ is the distance between a group head and the BS. The shadowing effect follows log-normal distribution with a variance of $\sigma^{2}_{s}$.

In practical applications, we aim at a target performance of $pM$ and $pF$, namely $pM_{obj}$ and $pF_{obj}$, respectively. Then, by substituting the target $pM_{obj}$ and $pF_{obj}$ into \eqref{neweq5}, the appropriate decision threshold $\theta$ and the required variance of $t^{th}$ AMP estimation $\tau_t$ can be determined while given the large-scale fading model. The associated solutions of $\theta$ and $\tau_t$ could be denoted as $\theta_{obj}$ and $\tau_{obj}$, respectively.

Bear the above statements in mind, in order to obtain $\theta_{obj}$ and $\tau_{obj}$, we shall specify the large-scale fading model. According to the proposed layered grouping network framework shown in Fig.\ref{fig1SystemModle}, the users of a cluster uniformly locates in the same ring, whose inner and outer radius are represented by $R_1$ and $R_2$, respectively. Hence the distance between a user and the BS obeys $d \sim [R_1,R_2]$. Accordingly, the probability density function of large-scale fading coefficient could be formulated as
\begin{equation}\label{neweq8}
P_{G}(g) \triangleq a_1g^{-\gamma_1}Q_1(g)-a_2g^{-\gamma_2}Q_2(g),
\end{equation}
where
\begin{equation}\label{neweq9}
\left\{
    \begin{array}{lr}
           a_{1} = \frac {40} {(R_2-R_1)^2\beta \sqrt{\pi}} exp(\frac{2(\ln 10)^2\sigma_s^2}{\beta^2} - \frac{2 \ln(10)\alpha}{\beta}),\\
		   a_{2} = \frac {40R1} {(R_2-R_1)^2\beta \sqrt{\pi}} exp(\frac{(\ln 10)^2\sigma_s^2}{2\beta^2} - \frac{\ln(10)\alpha}{\beta}),\\
		   \gamma_{1} \triangleq \frac {40} {\beta} +1,  \gamma_{2}\triangleq \frac {20} {\beta} +1,\\
           Q_i(g) = \int_{b\ln g+c_{i1}}^{b\ln g+c_{i2}} exp(-s^2)ds, i\in \{{1,2}\}\\
           c_{i2} = \frac{-\alpha - \beta log_{10}(R_1)} {\sqrt{2}\sigma_s} - \frac{20}{i \beta b}, i\in \{{1,2}\}\\
           c_{i1} = \frac{-\alpha - \beta log_{10}(R_2)} {\sqrt{2}\sigma_s} - \frac{20}{i \beta b}, i\in \{{1,2}\}\\
           b = -\frac {10\sqrt{2}}{\ln (10)\sigma_s}.\\
        \end{array}
\right.
\end{equation}

According to the state evolution method, if the AMP algorithm always achieves the target performance of $pM_{obj}$ and $pF_{obj}$, then the following inequality has to be true as long as a sufficient large iteration number $t$ is chosen
\begin{equation}\label{taut3}
\tau_{t+1} \leq \tau_{obj} \leq \tau_t.
\end{equation}
By substituting \eqref{taut2} into the above inequality, it results in
\begin{equation}\label{Lk2}
L_k \geq \frac {\frac {\sigma^2} {P_k}+ {M^{(k)}}MSE(\tau_t)} {\tau_{obj}^2}.
\end{equation}
Then, it is provable that $\mathrm{MSE \left( \cdot \right)}$ is a monotonically increasing function in the region of $g\in\left[0, 100\right]$. Hence we have $MSE(\tau_{obj}) \leq MSE(\tau_t)$ in practical scenarios. It implies replacing $MSE(\tau_t)$ by $MSE(\tau_{obj})$ in \eqref{Lk2} will yield a relaxed lower bound (LB) on MPL, which is given by
\begin{equation}\label{Lk3}
L_k \geq \frac {\frac {\sigma^2} {P_k}+ {M^{(k)}}MSE(\tau_{obj})} {\tau_{obj}^2}.
\end{equation}

For example, in practice, simultaneously achieving $pM_{obj}=0.05$ and $pF_{obj}=0.05$ may be an acceptable active group detection performance. While considering the large-scale fading model given in \eqref{neweq8}-\eqref{neweq9}, the associated solution of \eqref{neweq5} is $\theta_{obj} \approx 8.65 \times 10^{-8}, \tau_{obj} \approx 5 \times 10^{-8} $. Assuming that network configurations including the SNR, the cluster size $M^{(k)}$, as well as the sparsity $\lambda$ are known. Then, by substituting $\tau_{obj} \approx 5 \times 10^{-8} $ into \eqref{Lk3}, we could calculate the exact lower bound on MPL that enables the AMP algorithm to achieve the target active user detection performance.
\begin{figure}[htbp]
	\begin{center}
		\includegraphics[width=0.5\textwidth]{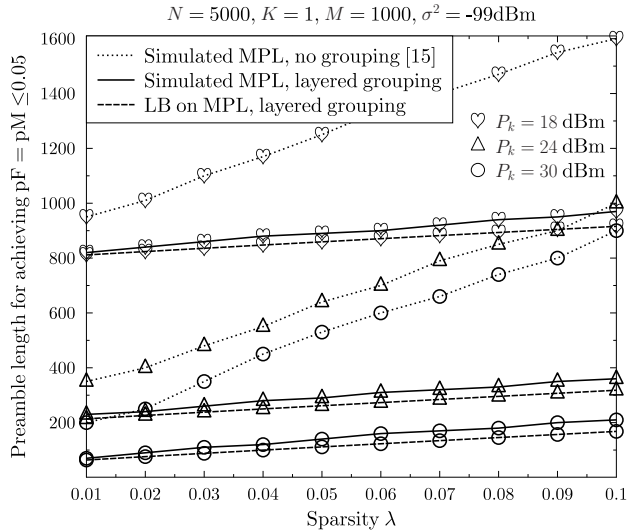}
		\caption{Comparison between simulated MPL and its lower bound with respect to different transmit powers. The practical preamble length required by conventional ``no grouping'' RA scheme \cite{Chen2018Sparse} is provided for comparison.}
		\label{fig4Bound1}
		\vspace{-0.1cm}
	\end{center}
\end{figure}

\begin{figure}[htbp]
	\begin{center}
		\includegraphics[width=0.5\textwidth]{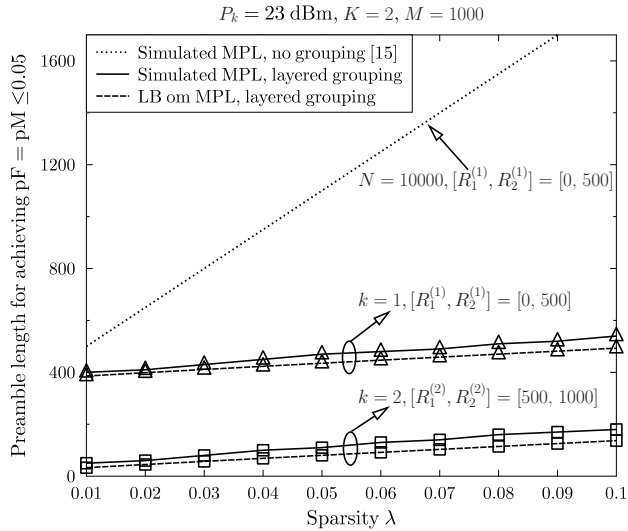}
		\caption{Comparison between simulated MPL and its lower bound with respect to different cluster coverages. The practical preamble length required by conventional ``no grouping'' RA scheme \cite{Chen2018Sparse} is provided for comparison.}
		\label{fig5Bound2}
		\vspace{-0.1cm}
	\end{center}
\end{figure}

It is evidenced in Fig.\ref{fig4Bound1} that the lower bounds on MPL \eqref{Lk3} gets quite close to the actual MPLs estimated by the Monte Carlo simulations for different transmit powers, although their discrepancy will be slightly enlarged while increasing the sparsity $\lambda$.
The comparison between simulated MPL and its lower bound with respect to different coverage areas are illustrated in Fig.\ref{fig5Bound2}. Fig.\ref{fig5Bound2} demonstrates that the discrepancy between the simulated MPL and its lower bound will be impacted by different cluster coverages. This phenomenon is due to the fact that the large-scale fading effect will be impacted by the cluster coverage as formulated in \eqref{neweq8} and \eqref{neweq9}. On the condition of having a low sparsity of $\lambda \leq 0.05$, this discrepancy would not exceed $10\%$ of the theoretical lower bound. Therefore, in algorithm \ref{algD}, we first calculate the lower bound on MPL for a specific cluster $c_k$, then the $L_k$ employed by the active groups in $c_k$ will be $10\%$ higher than the lower bound.

Furthermore, the conventional no grouping GF RA scheme proposed in \cite{Chen2018Sparse}, which also employs the MMSE based AMP algorithm, is shown in Fig.\ref{fig4Bound1} and Fig.\ref{fig5Bound2} as well. It is demonstrated in Fig.\ref{fig4Bound1} and Fig.\ref{fig5Bound2} that the MPL required by the proposed layered grouping based RA scheme is always significantly less than that required by its counterpart in \cite{Chen2018Sparse} regardless of different SNR values, different sparsities, different coverages.  For the sake of fair comparison, the total number of cellular users remains the same in Fig.\ref{fig4Bound1} and Fig.\ref{fig5Bound2}.

\section{Simulation Results and Discussions}
\label{sec:simulation}
In this section, the active user detection performance of the proposed \textcolor[rgb]{0,0,0}{two-phase DPS aided} RA scheme is simulated. \textcolor[rgb]{0,0,0}{The obtained results are} compared with the \textcolor[rgb]{0,0,0}{classical CS aided RA} that does not exploit any grouping strategy\textcolor[rgb]{0,0,0}{\cite{Chen2018Sparse} and with the conventional group paging aided RA that does not exploit CS technologies\cite{2016Group}.} Without loss of generality, the group size $|g_{k,m}|$ \textcolor[rgb]{0,0,0}{and the} number of groups in different clusters $M^{(k)}$ \textcolor[rgb]{0,0,0}{are prefixed to constants} regardless of the \textcolor[rgb]{0,0,0}{indices of $k$ and $m$.} Other system parameters are listed in Table.\ref{t2}.
 \begin{table}
 \begin{center}
 \small
 \setlength{\tabcolsep}{0.01mm}{
 \caption{System Configuration.}
 \begin{tabular}{c|c}
 \hline
  Parameter    &   Value    \\ \hline
  Radius of the cell   & $1000 m$ \\
  Path-loss model       & $15.3+37.6\log_{10}(d(m))$    \\
  Variance of shadowing $\sigma_s^2 $  & $8$    \\
  Background noise $N_o$      & -99dBm    \\
  Total cellular devices $N$       & $10000, 20000 $   \\
  Number of clusters $K$		& \textcolor[rgb]{0,0,0}{$2, 4, 8$}		\\
  Group size $\vert g_{k,m} \vert$      & $5$   \\
  Number of groups per cluster $M^{(k)}$  &  \textcolor[rgb]{0,0,0}{$M= 1000, 2000$} \\
  Preamble type and length of $\mathbf{s}^{I}_{k}$ 	&   Walsh Seq., $\vert\mathbf{s}^{I}_{k} \vert =  32$\\
  TX power $P$ defined in \eqref{rho}	&  23dBm \\
  Preamble type of $\mathbf{s}^{II}_{k,m}$  &  Gaussian Random Seq.\\
  TX power $P_k$ defined in \eqref{yg}	&  23dBm \\   \\
  Cluster coverages of $K=2$& $\left[ {{R_1^{(k)}},{R_2^{(k)}}} \right] = \left\{ {\begin{array}{*{20}{c}}
  	{\left[ {0,500} \right]\;,\;\;\;k = 1}\\
  	{\left[ {500,1000} \right],\;\;\;k = 2}
  	\end{array}} \right.$	\\ \\
  Cluster coverages of $K=4$& $\left[ {{R_1^{(k)}},{R_2^{(k)}}} \right] = \left\{ {\begin{array}{*{20}{c}}
  	{\left[ {0,250} \right]\;,\;\;\;k = 1}\\
  	{\left[ {250,500} \right],\;\;\;k = 2}\\
  	{\left[ {500,750} \right],\;\;\;k = 3}\\
  	{\left[ {750,1000} \right],\;\;\;k = 4}\\
  	\end{array}} \right.$	\\	\hline
 \end{tabular}
 \label{t2}}
 \end{center}
 \end{table}

\begin{figure}[htbp]
	\begin{center}
		\includegraphics[width=0.55\textwidth]{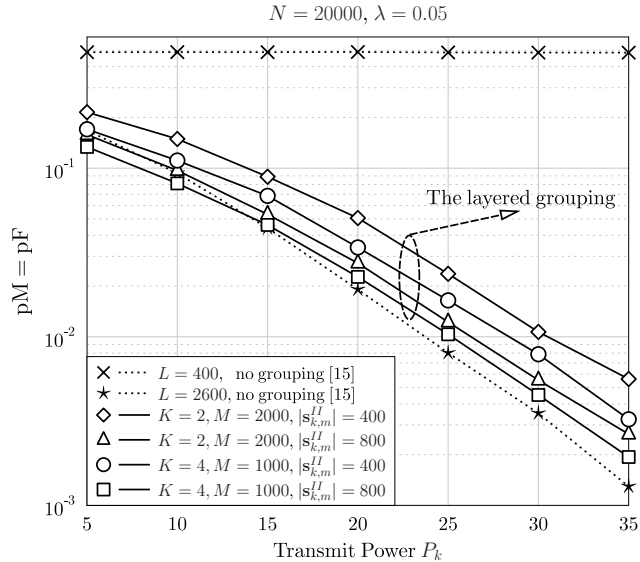}
		\caption{$\mathrm{pF}$, $\mathrm{pM}$ versus transmit power of \textcolor[rgb]{0,0,0}{the two-phase DPS aided RA}. It is clear that the layered grouping scheme employing a preamble length of only 800 can approach the performance of no grouping scheme \cite{Chen2018Sparse} which employs a very long preamble length of 2600.}
		\label{fig6pFpM}
		\vspace{-0.1cm}
	\end{center}
\end{figure}

In Fig.\ref{fig6pFpM}, the $pF$, $pM$ versus transmit power in the \textcolor[rgb]{0,0,0}{two-phase DPS aided} RA is compared with that of \textcolor[rgb]{0,0,0}{CS aided RA} \cite{Chen2018Sparse}, which also employs the MMSE based AMP algorithm. It is a general consensus that $pF=pM$ \textcolor[rgb]{0,0,0}{implies a good performance balance of active user detection can be achieved.} Hence the decision threshold employed by \textcolor[rgb]{0,0,0}{MMSE based AMP} algorithm is adjusted for achieving $pF=pM$.

Comparing the solid line labelled by diamonds with the dashed line labelled by crosses in Fig.\ref{fig6pFpM}, the \textcolor[rgb]{0,0,0}{two-phase DPS aided RA} achieves a dramatic power gain with respect to the conventional \textcolor[rgb]{0,0,0}{CS based RA} \cite{Chen2018Sparse} while using the same preamble length of 400. Comparing the solid line labelled by triangles with the solid line labelled by squares in Fig.\ref{fig6pFpM}, the grouping strategy of $K=4,\ M=1000$  achieves a better performance than that of $K=2,\ M=2000$, even they employ the same preamble length of 800. It implies a \textcolor[rgb]{0,0,0}{RA power} gain is available by adjusting the number of clusters and the number of groups.

Then, we define the probability of successful detection ($pS$) in cluster $c_k$ as
\begin{equation}\label{pS}
	pS^{(k)}=\frac{\sum_{m=1}^{M^{(k)}}\{\hat{x}_{k,m} \geq \theta \ \&\ x_{k,m}\neq 0 \}}{\sum_{m=1}^{M^{(k)}}\{x_{k,m}\neq 0\}}.
\end{equation}
According to this definition and equation \eqref{neweq5}, $pF$, $pM$ and $pS$ have following relationships

\begin{equation}\label{neweq12}
\left\{
    \begin{array}{l}
           pS^{(k)} = 1-pM^{(k)} = 1-\int (1-e^{-\frac{\theta^2}{\tau_t^2+g^2}})\cdot P_G^{(k)}(g)dg,\\
           pS^{(k)} = 1-\frac {1-\lambda_k} {\lambda_k}pF^{(k)} =1-\frac {1-\lambda_k} {\lambda_k}e^{-\frac{\theta^2}{\tau_t^2}}.
        \end{array}
\right.
\end{equation}

The parameter $\tau_t$ involved in \eqref{neweq12} can be determined according to \eqref{taut2}. \textcolor[rgb]{0,0,0}{Hence, \eqref{neweq12} enable us to theoretically analyze the active group detection performance of the proposed RA strategy.} 

\begin{figure}[htbp]
	\begin{center}
		\includegraphics[width=0.5\textwidth]{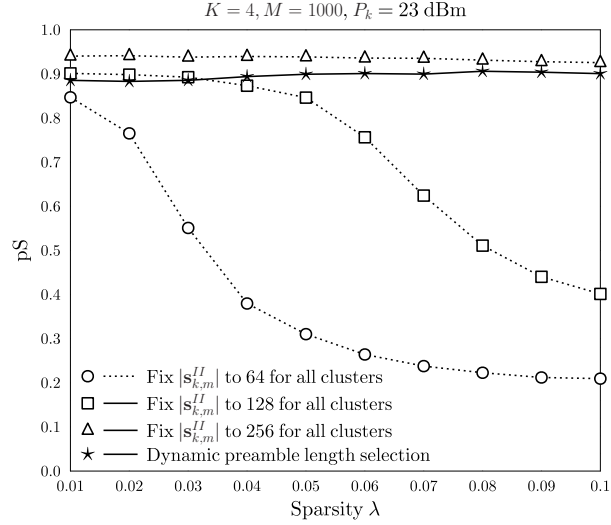}
		\caption{ $\mathrm{pS}$ versus sparsity performance of two-phase \textcolor[rgb]{0,0,0}{DPS aided} RA, where preamble length fixed strategy \textcolor[rgb]{0,0,0}{is compared with.} The preamble length required by dynamic scheme is always less than 210 after averaged over all clusters.}
		\label{fig8Dynamic}
		\vspace{-0.1cm}
	\end{center}
\end{figure}

\textcolor[rgb]{0,0,0}{As a result, the} $pS$ achieved by preamble length fixed strategy is compared with that achieved by \textcolor[rgb]{0,0,0}{DPS} strategy in Fig.\ref{fig8Dynamic}, \textcolor[rgb]{0,0,0}{where both of them employ two-phase RA framework and the grouping strategy is fixed to $K=4, M= 1000$.}
Observe at Fig.\ref{fig8Dynamic} that while employing a predefined preamble length of $|\mathbf{s}_{k,m}^{II}|=64,\ 128$, the $pS$ of \textcolor[rgb]{0,0,0}{grouped} RA still rapidly drops along with the growth of sparsity. In contrast, benefiting from the \textcolor[rgb]{0,0,0}{DPS aided RA} scheme, the \textcolor[rgb]{0,0,0}{system} is capable of achieving a high $pS$ probability throughout the entire sparsity region. Actually, the \textcolor[rgb]{0,0,0}{ DPS strategy} approaches a similar active group detection performance to a preamble length fixed counterpart having $\vert \mathbf{s}^{II}_{k,m} \vert = 256$. However, the average preamble length required by the \textcolor[rgb]{0,0,0}{DPS} strategy is always less than 210 throughout the entire sparsity region.  It \textcolor[rgb]{0,0,0}{means} the \textcolor[rgb]{0,0,0}{DPS} strategy will \textcolor[rgb]{0,0,0}{further save} considerable overhead.



We would like to further evaluate the energy \textcolor[rgb]{0,0,0}{consumption} of two-phase \textcolor[rgb]{0,0,0}{DPS aided} RA. According to interpretation in Section \ref{sec:System Model}, the transmission energy required for constructing the layered grouping network framework is negligible. Because the group construction only happens once after a GH user is registered. Besides, the group update only happens once during a very long period. 
\textcolor[rgb]{0,0,0}{Hence we only focus on the random access energy desired on the active user side in the entire “Phase-I” plus “Phase-II” durations.  As shown in Fig.\ref{fig3RandomAccess},  it consists of six parts: (1) energy required by transmitting cluster preambles for the sake of cluster load estimation, i.e. $\epsilon_{\mathrm{pmb-I}}=\frac 1 {N\cdot\lambda} {\sum_{k=1}^{K}\sum_{m=1}^{M^{(k)}}a_{k,m}P_{k,m}|\mathbf{s}_k^I|}$, (2) energy consumed by waiting for the Phase-II solution message, i.e. $\epsilon_{\rm{wait-I}}= P_{\rm{wait}}\cdot \overline{T}_{\rm{wait-I}}$, (3) energy required by processing the received Phase-II solution message, i.e. $\epsilon_{\mathrm{pcs-I}}= P_{\rm{pcs}}\cdot \overline{T}_{\rm{pcs-I}}$,  (4) energy required by transmitting access preambles for the sake of active group identification, i.e. $\epsilon_{\mathrm{pmb-II}}=\frac 1 {N\cdot\lambda} {\sum_{k=1}^{K}\sum_{m=1}^{M^{(k)}}a_{k,m}P_{k,m}|\mathbf{s}_k^{II}|}$,  (5) energy consumed by the waiting model (as defined in \cite{Sesia2009LTE}) in the entire “Phase-II” duration, i.e. $\epsilon_{\rm{wait-II}}= P_{\rm{wait}}\cdot \overline{T}_{\rm{wait-II}}$, (6) energy required by processing the PDTS message, i.e. $\epsilon_{\mathrm{pcs-II}}= P_{\mathrm{pcs-II}}\cdot \overline{T}_{\mathrm{pcs-II}}$. Hence, the average random access energy per active user  could be calculated as:}

\begin{equation}\label{energyTotal}
	\textcolor[rgb]{0,0,0}{\overline{\epsilon} = \epsilon_{\rm{pmb-I}} + \epsilon_{\rm{wait-I}} + \epsilon_{\rm{pcs-I}} + \epsilon_{\rm{pmb-II}} + \epsilon_{\rm{wait-II}} + \epsilon_{\rm{pcs-II}}}
\end{equation}
\textcolor[rgb]{0,0,0}{In the above energy parts, $P_{k,m}$ and $P_k$ are defined in \eqref{rho} and \eqref{yg}, respectively. Thus, the values of $P_{\rm{wait}}$, $P_{\rm{pcs}}$, $\overline{T}_{\rm{wait-I}}$ and $\overline{T}_{\rm{wait-II}}$ are specified according to similar parameters given in \cite{2016Group}-\cite{Sesia2009LTE}. Particularly, $\overline{T}_{\rm{wait-II}}$ is the average waiting time of an active group required in the entire “Phase-II” duration, which equals to subtracting length of $|\mathbf{s}_k^{II}|$ and $\overline{T}_{\rm{pcs-II}}$ from duration of “Phase-II” \footnote{The duration of “Phase-II” is determined by system configurations of $N$,$K$,$\lambda$ etc. It can be calculated by experimental method as shown in Fig.\ref{fig12Delay}.}. More specifically, according to LTE standard, up to 839 symbols can be transmitted within a single time slot (i.e. 0.5 ms). Hence, we equivalently employ the number of symbols as our time metric.}

\textcolor[rgb]{0,0,0}{Furthermore, if we replace practical preamble length $|\mathbf{s}_k^{II}|$ used in \eqref{energyTotal} by its theoretical lower bound (LB) specified in \eqref{Lk3}, the LB of $\epsilon_{\rm{pmb-II}}$ is given by} 
\begin{equation}\label{energypreambleII}
	\textcolor[rgb]{0,0,0}{\epsilon_{\rm{pmb-II}}^* = \frac 1 {N\cdot\lambda} {\sum_{k=1}^{K}\sum_{m=1}^{M^{(k)}}a_{k,m}P_{k}\frac {\frac {\sigma^2} {P_k}+ {M^{(k)}}MSE(\tau_{obj})} {\tau_{obj}^2}}},
\end{equation}
\textcolor[rgb]{0,0,0}{Simultaneously, the length of $\overline{T}_{\rm{wait-II}}$ is also minimized, which results in the minimization of $\epsilon_{\rm{wait-II}}$, namely $\epsilon_{\rm{wait-II}}^*$. Substitute $\epsilon_{\rm{pmb-II}}^*$ and $\epsilon_{\rm{wait-II}}^*$ into \eqref{energyTotal}, we refer to the resultant $\overline{\epsilon}$ as theoretical RA energy of our two-phase DPS aided RA, namely $\overline{\epsilon}^*$.}

\begin{figure}[htbp]
	\begin{center}
		\includegraphics[width=0.5\textwidth]{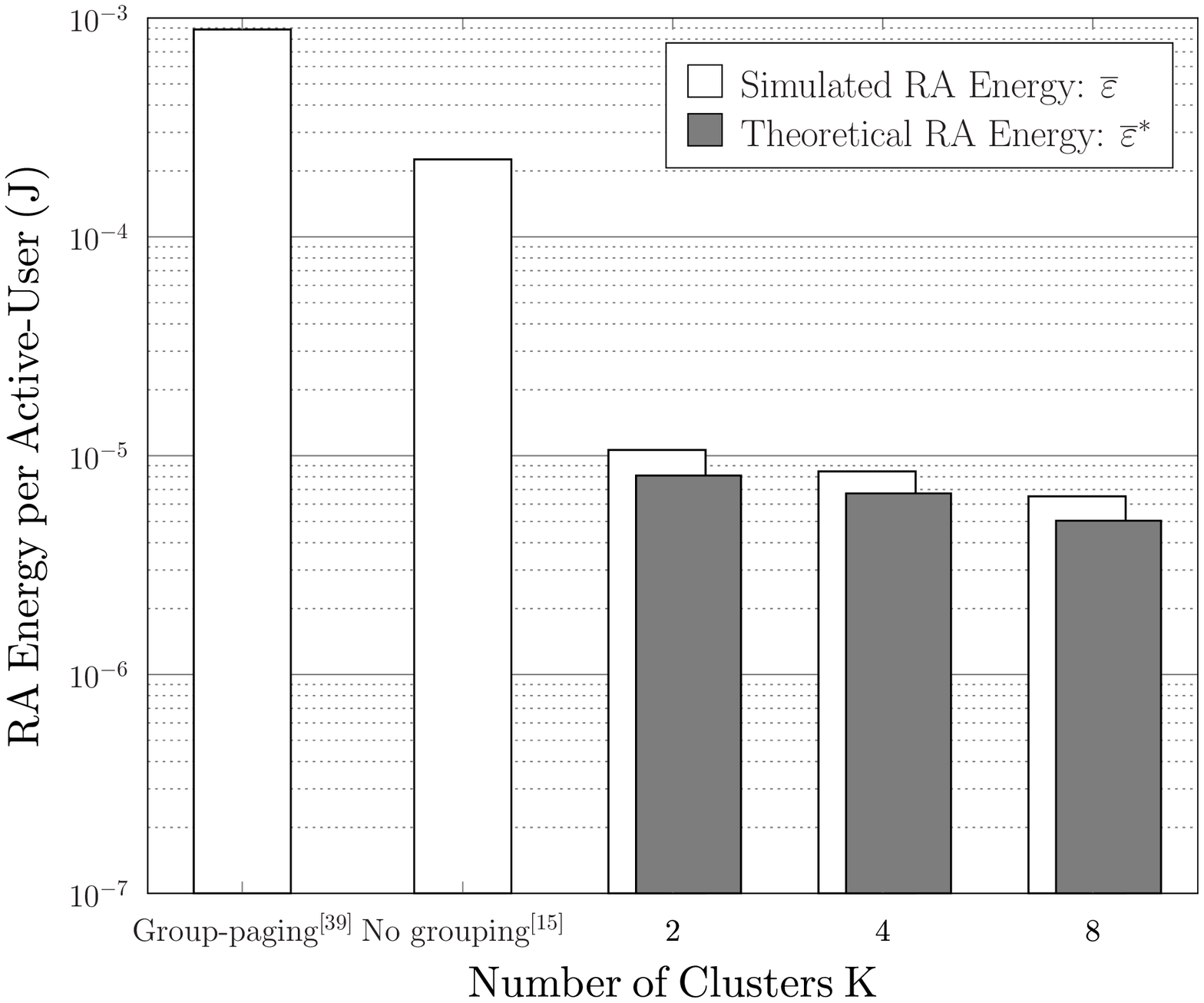}
		\caption{\textcolor[rgb]{0,0,0}{RA energy comparison between the proposed scheme, group-paging RA scheme\cite{2016Group} and no grouping RA scheme\cite{Chen2018Sparse}, where $N =20000$, $\lambda = 0.05$, $K$ increases from 2 to 8.}}
		\label{fig9EC}
		\vspace{-0.1cm}
	\end{center}
\end{figure}

\begin{figure}[htbp]
	\begin{center}
		\includegraphics[width=0.5\textwidth]{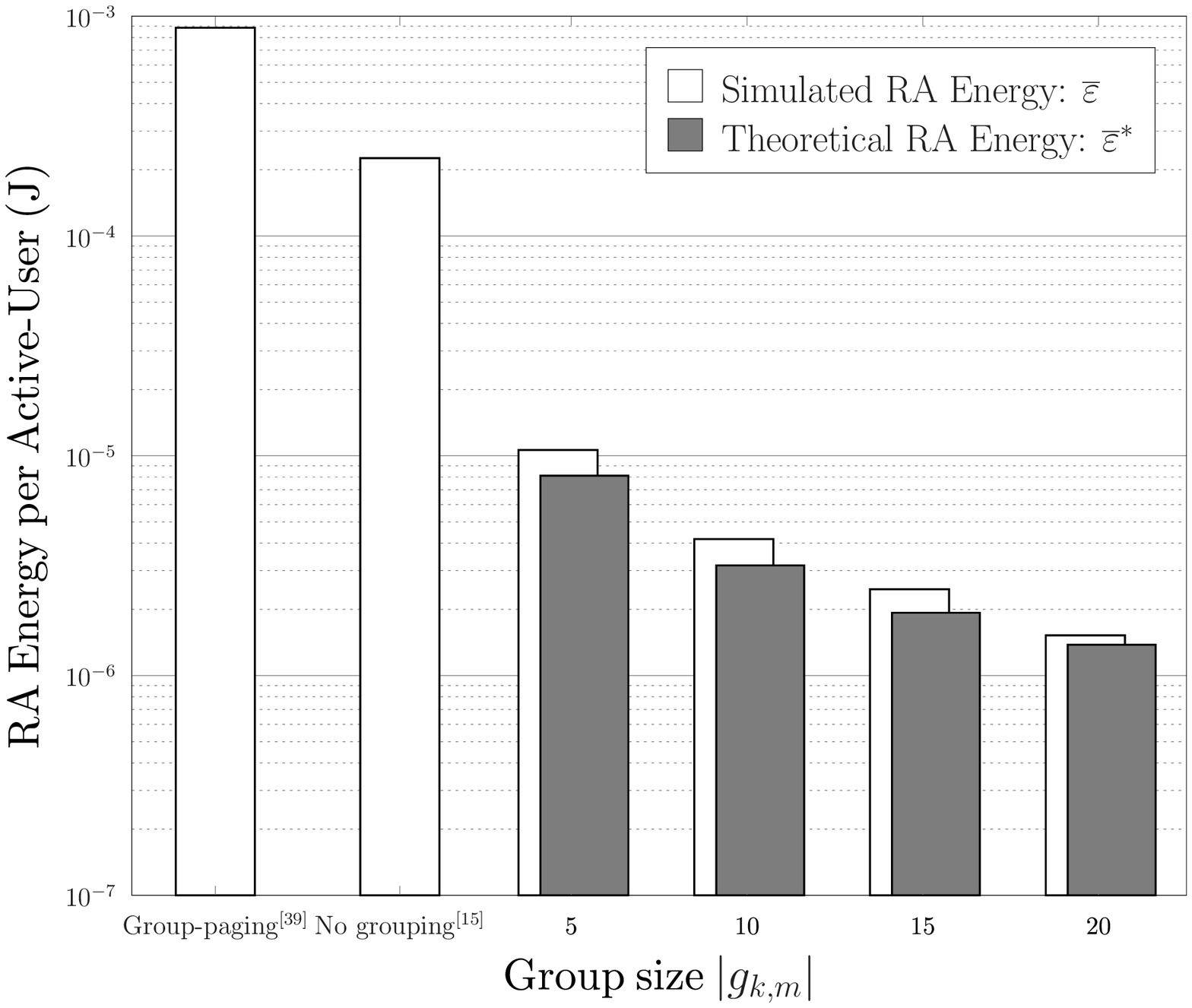}
		\caption{\textcolor[rgb]{0,0,0}{RA energy comparison between the proposed scheme, group-paging RA scheme \cite{2016Group} and no grouping RA scheme \cite{Chen2018Sparse}, where $N =20000$, $\lambda = 0.05$, $\vert g_{k,m} \vert$ increases from 5 to 20.}}
		\label{fig10ECgsize}
		\vspace{-0.1cm}
	\end{center}
\end{figure}

Observe at Fig.\ref{fig9EC} and Fig.\ref{fig10ECgsize} that on the condition of approaching the same $pS \geq 90\%$, the two-phase \textcolor[rgb]{0,0,0}{DPS aided} RA scheme achieves a significant power gain compared with \textcolor[rgb]{0,0,0}{both the} conventional no grouping \textcolor[rgb]{0,0,0}{CS based} RA scheme \cite{Chen2018Sparse} \textcolor[rgb]{0,0,0}{and the conventional no CS aided group paging strategy\cite{2016Group}.} \textcolor[rgb]{0,0,0}{There are three major advantages of the proposed layered grouping RA scheme, i.e. (1) in the proposed system, only GH of an active group has to send access preamble, while in \cite{Chen2018Sparse},\cite{2016Group}, every active user has to send access preamble; (2) GHs also have better channel conditions than other group members owing to the Opt-EC based K-means grouping algorithm; (3) the DPS algorithm given in Algorithm \ref{algD} effectively reduces the preamble length used in phase-II. In more details, the energy consumed per active user is reduced as the number of clusters K is increased, as illustrated in Fig.\ref{fig9EC}, while the associated penalty is the increase of RA delay as shown in Fig \ref{fig12Delay} per active user versus. On the other hand, as can be seen from Fig.\ref{fig10ECgsize}, the energy consumed per active user can be reduced more significantly when the group size is increased.  However, it may become impractical if the group size is too big due to the complexity and synchronization requirements.}

\begin{figure}[htbp]
	\begin{center}
		\includegraphics[width=0.5\textwidth]{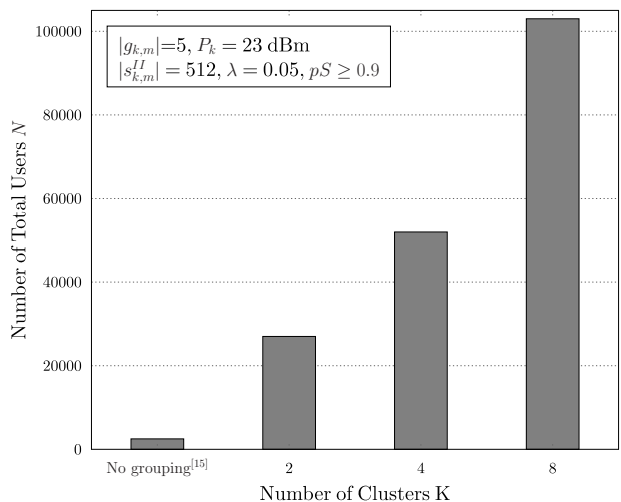}
		\caption{\textcolor[rgb]{0,0,0}{Comparison on affordable number of coexisted users in a cell between the proposed RA and no grouping CS aided RA.}}
		\label{fig11Capacity}
		\vspace{-0.1cm}
	\end{center}
\end{figure}

\begin{figure}[htbp]
	\begin{center}
		\includegraphics[width=0.5\textwidth]{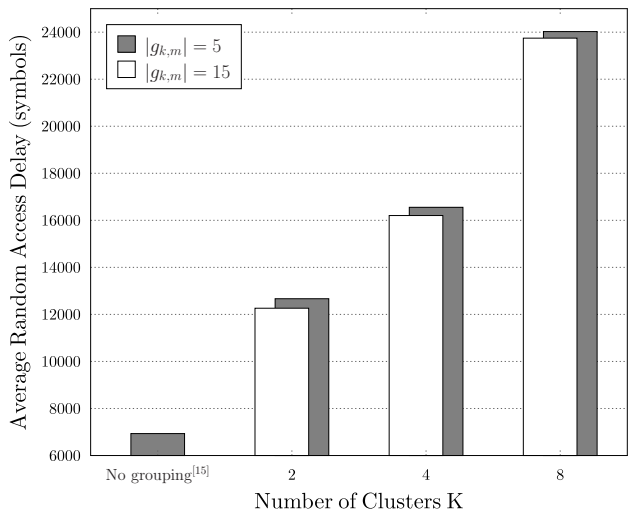}
		\caption{Access delay versus number of clusters in the proposed two-phase GF RA scheme.}
		\label{fig12Delay}
		\vspace{-0.1cm}
	\end{center}
\end{figure}

\textcolor[rgb]{0,0,0}{The capability of DPS aided RA scheme to support the massive connectivity is depicted in Fig. \ref{fig11Capacity}, which is measured in the maximum number of coexisted users in a cell while fixing the preamble length, the sparsity, the transmit power, as well as the target level of $pF$ and $pM$. As can be observed from Fig. \ref{fig11Capacity}, for the given condition, the proposed RA scheme is capable of supporting more than $10^{5}$ users. In contrast, no grouping strategy \cite{Chen2018Sparse} can only support} \textcolor[rgb]{0,0,0}{approximately 2500 }\textcolor[rgb]{0,0,0}{users for the same amount of physical resources.}

\textcolor[rgb]{0,0,0}{Finally,} the average \textcolor[rgb]{0,0,0}{time} required by an active group head $\dot{u}_{k,m}$ for completing random access procedure is adopted as our delay metric. According to the proposed \textcolor[rgb]{0,0,0}{RA} procedure in Fig.\ref{fig3RandomAccess} \textcolor[rgb]{0,0,0}{and in line with our RA energy analysis,} the \textcolor[rgb]{0,0,0}{RA} delay of $\dot{u}_{k,m}$ consists of \textcolor[rgb]{0,0,0}{six} major components:
(1) The time $\overline{T}_1$ required by all active GHs for transmitting their cluster preambles of $|\bm{s}_k^I|,\; k=1,2,\cdots, K$; \textcolor[rgb]{0,0,0}{(2) the waiting time $\overline{T}_{\rm{wait-I}}$ during "Phase-I";  (3) the time $\overline{T}_{\rm{pcs-I}}$ required for processing phase-II solution message; (4) the time $\overline{T}_{2}$ required by an active GH for transmitting its group access preamble; (5) the waiting time $\overline{T}_{\rm{wait-II}}$ during “Phase-II” and (6) the time $\overline{T}_{\rm{pcs-II}}$ required for processing PDTS message. Again, the number of symbols is employed as the time metric.}



The random access delay performance of \textcolor[rgb]{0,0,0}{the proposed DPS aided} RA scheme is demonstrated in Fig.\ref{fig12Delay}, where $\lambda = 0.05,\; N=20000$. Observe at Fig.\ref{fig12Delay} that the \textcolor[rgb]{0,0,0}{proposed} RA scheme imposes a higher random access delay than the conventional no grouping \textcolor[rgb]{0,0,0}{CS based RA scheme, especially when the number of clusters $K$ grows. But, benefiting from the DPS scheme, employing a large group size could slightly mitigate the latency.} Consequently, \textcolor[rgb]{0,0,0}{our proposal} may be more suitable for the \textcolor[rgb]{0,0,0}{mMTC devices which} have a relative higher tolerance of time delay. Fortunately, benefiting from the \textcolor[rgb]{0,0,0}{DPS aided RA} scheme, the random access delay will not linearly increase with respect to the number of clusters.

\section{Conclusions}
\label{sec:conclu}
With \textcolor[rgb]{0,0,0}{the} explosive growth of IoT \textcolor[rgb]{0,0,0}{devices (may approach around 125 billion by 2030 \cite{Sharma2020Toward}),} the mMTC communication will become one of the most important services of forthcoming B5G networks. An extremely large number of devices are accommodated in a single cell. Neighborhood devices are inclined to have a similar communication behavior and QoS requirement. These fundamental characteristics of future mMTC scenario motivate us to propose a two-phase \textcolor[rgb]{0,0,0}{DPS aided} RA scheme. \textcolor[rgb]{0,0,0}{Benefiting from the proposed layered grouping network framework,} instead of a large number of active users directly access the BS, only the GHs of active groups access the BS on behalf of all the active members. The mechanism of constructing and maintaining this layered grouping network framework, as well as the two-phase RA \textcolor[rgb]{0,0,0}{procedure} are carefully designed, especially the Opt-EC based K-means grouping algorithm, the orthogonal sequence based cluster load estimation, the dynamic preamble selection, as well as the MMSE based AMP algorithm. A tight lower bound on MPL required by AMP algorithm for achieving the given detection accuracy is provided. \textcolor[rgb]{0,0,0}{In summery, compared with the existing RA schemes, the proposed DPS aided RA scheme
 achieves three major improvements: a) reducing the access overhead as shown in Fig. \ref{fig4Bound1}, Fig. \ref{fig5Bound2} and Fig.\ref{fig8Dynamic}; b) saving the access energy as shown in Fig.\ref{fig6pFpM} and Fig.\ref{fig9EC}; c) increasing the number of coexisted cellular users as shown inFig.\ref{fig11Capacity}, at the price of relatively longer access delay.}

{
\bibliographystyle{ieeetr}
\bibliography{refrence}
}

\end{document}